# Perpendicular magnetic anisotropy of full-Heusler films in $Pt/Co_2FeAl/MgO$ trilayers

Xiaoqi Li, Shaoqian Yin, Yupeng Liu, Delin Zhang, Xiaoguang Xu *, Jun Miao, and Yong Jiang *

State Key Laboratory for Advanced Metals and Materials, School of Materials Science and Engineering, University of Science and Technology Beijing, Beijing 100083, China.

We report on perpendicular magnetic anisotropy (PMA) in a $Pt/Co_2FeAl/MgO$ sandwiched structure with a thick $Co_2FeAl$ layer of 2-2.5 nm. The PMA is thermally stable that the anisotropy energy density $K_u$ is $1.3\times10^6$ erg/cm$^3$ for the structure with 2 nm $Co_2FeAl$ after annealing at 350 $^oC$. The thicknesses of $Co_2FeAl$ and MgO layers greatly affect the PMA. Our results provide an effective way to realize relative thick perpendicularly magnetized Heusler alloy films.

* Email: xgxu@ustb.edu.cn, yjiang@ustb.edu.cn

Magnetic tunnel junctions (MTJs) as core elements of magnetic random access memory (MRAM) have gotten aroused enormous interests for their large magnetoresistance (MR). Because the large MR comes from spin-polarized tunneling, [1, 2] spin polarization at Fermi level ($E_F$) of the two ferromagnetic (FM) electrodes in MTJs is a key factor to determine MR based on Julliere's model. [3] Therefore half-metals, which have 100% theoretical spin polarization, are ideal electrodes for MTJs. [4-6]

A prediction shows that FM electrodes with perpendicular magnetic anisotropy (PMA) could bring faster and smaller MTJs than those with in-plane magnetic anisotropy. [7] However till now no traditional PMA film can satisfy the requirements of high thermal stability at reduced dimension, low critical current for spin-torque-induced switching and high MR simultaneously. [8, 9] Recently, Pt (or Pd)/Co (or CoFeB)/$MO_x$ (M=Mg and Al, etc.) trilayers with strong PMA were reported. [10-15] The PMA was supposed to origin from the hybridization of Co 3d and O 2p orbitals. A high MR ratio of 120% was also obtained in perpendicular CoFeB/MgO/CoFeB MTJ. [9] Due to the high spin polarization, it should be a big breakthrough if one can realize half metals with PMA. Perpendicularly magnetized Pt/$Co_2FeAl$ has really been fabricated. [16] However the thin thickness of 0.6 nm may cause the deterioration of spin polarization and thermal stability for $Co_2FeAl$ films. In this letter, we report thick $Co_2FeAl$ films of 2-2.5 nm with PMA in Pt/$Co_2FeAl$/MgO trilayers. The perpendicular $Co_2FeAl$ films show good thermal stability after 350 $^{o}$C annealing.

All the thin films were deposited on $Si/SiO_2$ substrates with a buffer Ta layer of 3 nm by magnetron sputtering. The stack structure is $Si/SiO_2$/Ta (3)/Pt (20)/$Co_2FeAl$ ($t_{CFA}$)/MgO ($t_{MgO}$), (layer thickness in nanometer, $t_{CFA}$ and $t_{MgO}$ represent the thicknesses of $Co_2FeAl$ and MgO, respectively). The base pressure of the sputtering system was better than $9.0\times10^{-6}$ Pa. The annealing process was carried out in a vacuum chamber under $2\times10^{-4}$ Pa for 0.5 hour in the absence of magnetic field. The thin films were then characterized by alternating gradient magnetometer (AGM), magnetic force microscopy (MFM), and X-ray photoelectron spectroscopy (XPS), respectively.

Figure 1 shows the M-H loops for the as-deposited and annealed samples with $t_{CFA}$= 2 nm and $t_{MgO}$= 0.5 nm. No apparent PMA is observed for the as-deposited film. After annealing at 150 $^{o}$C, the PMA is clearly realized. It is similar to Pd (or Pt)/CoFeB (or Co)/MgO structure in which there is also a transition from in-plane magnetic anisotropy to PMA after annealing, which was due to the oxidation of CoFeB (or Co). [14, 15] For our samples, while the annealing temperature ($T_{an}$) is as high as 350 $^{o}$C, the PMA is stably maintained and the anisotropy energy density $K_u$ is estimated to be about $1.3\times10^6$ erg/cm$^3$ by using the equation $K_u=M_sH_{an}/2$, where $M_s$ and $H_{an}$ refer to saturation magnetization and anisotropy field, respectively. Therefore the PMA in the structure has a good thermal stability.

The perpendicular coercivity ($H_c$) and saturation field ($H_s$) as functions of $T_{an}$ are plotted in Figure 2. $H_s$ decreases sharply from 6000 Oe for the as-deposited sample down to 350 Oe for the 150 $^{o}$C annealed one. At the same time, $H_c$ increases from 18 to 40 Oe. When $T_{an}$ further increases up to 350 $^{o}$C, $H_s$ only has a slight change and $H_c$ reaches a maximum value (60 Oe). The inset of Fig. 2 shows the MFM image of the sample annealed at 350 $^{o}$C. The regularly spaced stripe domains further justify the existence of PMA. When the sample is annealed at 400 $^{o}$C, $H_s$ increases to 7600 Oe. The PMA disappears. So the annealing process plays a crucial role for the PMA, which is different from Pt/$Co_2FeAl$ multilayers. [16]

Because one possible origin of the PMA is the hybridization of Co 3d and O 2p orbitals, [12] XPS is performed for the sample of Pt (20)/$Co_2FeAl$ (2)/MgO (0.5) annealed at 350 $^{o}$C to obtain any information on the composition of Co and Fe. The Co and Fe 2p spectra are shown in Figure 3. The main peaks of CoO $2p_{3/2}$, CoO $2p_{1/2}$ for Co 2p and $Fe_2O_3$ $2p_{3/2}$, $Fe_2O_3$ $2p_{1/2}$ for Fe 2p have been detected, demonstrating that the $Co_2FeAl$ film is partially oxidized. As a result, Co-O and Fe-O bonds are coexisting. Besides, two satellite peaks of Co 2p (marked as 'S') can be distinguished which arise from the charge transfer between Co 3d and O 2p. [17-19] There is a similar behavior between Fe 3d and O 2p. The charge transfer at the $Co_2FeAl$/MgO interface may create a strong band splitting and lead to the PMA. [12]

Figure 4 shows $M_s$ and the effective thickness of $Co_2FeAl$ ($t_{eff}$) as a function of $t_{MgO}$ for the as-deposited Pt (20)/$Co_2FeAl$ (2)/MgO ($t_{MgO}$) samples. The values of $t_{eff}$ are estimated by comparing $M_s$ of each sample with that of the Pt (5)/$Co_2FeAl$ (2)/Pt (4). It is unexpected that $M_s$ monotonously decreases with the increasing $t_{MgO}$. The strong dependence of $M_s$ on the thickness of oxide layer has not ever been mentioned in Pt/Co (or CoFeB)/$MO_x$ systems, [12-15] and is due to different degree of oxidation at the $Co_2FeAl$/MgO interface. It is reasonably assumed that the thicker sputtered MgO film leads to a deeper oxidation at the interface. There is also an optimum degree of oxidation leading to PMA in the Pt/$Co_2FeAl$/MgO structure, which is similar to Pt/Co/$MO_x$ trilayers. [12] $t_{eff}$ is 1.3 nm in the as-deposited state for the sample with $t_{CFA}$=2 nm and $t_{MgO}$=0.5 nm. After annealing at 150 $^oC$, $t_{eff}$ decreases to 1.1 nm and the sample shows PMA. On the other hand, the Pt (5)/$Co_2FeAl$ (2)/Pt (4) sample keeps in-plane anisotropy in both as-deposited and annealed states. Therefore the oxidation at the $Co_2FeAl$/MgO interface plays a crucial role in the PMA.

While $t_{CFA}$= 2.5 nm and $t_{MgO}$= 0.5 nm, we have also deposited samples with the structure of Pt (20)/$Co_2FeAl$ (2.5)/MgO (0.5). The as-deposited sample is in-plane magnetized, as shown in the inset of Fig. 5. After annealing at 250 $^oC$, the sample becomes PMA as shown in Fig. 5. The PMA is maintained after 350 $^oC$ annealing, as in the inset of Fig. 6. Fig. 6 plots the variation of the perpendicular $H_s$ for the samples Pt (20)/$Co_2FeAl$ ($t_{CFA}$)/MgO (0.5) with different $t_{CFA}$ in both as-deposited and 350 $^oC$ annealed states, respectively. All the samples annealed at 350 $^oC$ have lower $H_s$ compared with the as-deposited state because more Co-O and Fe-O bonds are created to enhance the PMA after annealing. Large $H_s$ values of the samples with $t_{CFA}$<2 nm demonstrate their in-plane anisotropy. It may be due to the excessive oxidation leading to a too small $t_{eff}$. The minimum value of $H_s$ apears in the 350 $^oC$ annealed sample with $t_{CFA}$=2 nm.

In conclusion, the PMA is observed in the trilayers Pt/$Co_2FeAl$ /MgO while the thicknesses of the $Co_2FeAl$ film are 2 and 2.5 nm. The PMA can be stablely maintained after 350 $^{o}$C annealing. A possible origin of the PMA is the Co-O and Fe-O bands hybridization. Therefore we suppose that all Co-based Heusler alloy films with PMA can be realized in the structure of Pt/*Heusler alloy*/MgO.

**Acknowledgements**

This work was partially supported by the NSFC (Grant Nos. 50831002, 50971025, 51071022), the Keygrant Project of Chinese Ministry of Education (No. 309006) and the National Basic Research Program of China (Grant No. 2007CB936202).

## References

[1] J. S. Moodera, L. R. Kinder, T. M. Wong, and R. Meservey: Phys. Rev. Lett. **74** (1995) 3273.

[2] T. Miyazaki, and N. Tezuka: J. Magn. Magn. Mater. **139** (1995) L231.

[3] M. Julliere: Phys. Lett. A **54** (1975) 225.

[4] W. Wang, H. Sukegawa, R. Shan, S. Mitani, and K. Inomata: Appl. Phys. Lett. **86** (2005) 232503.

[5] S. Okamura, A. Miyazaki, S. Sugimoto, N. Tezuka, and K. Inomata: Appl. Phys. Lett. **86** (2005) 232503.

[6] T. Marukame, T. Ishikawa, S. Hakamata, K. Matsuda, T. Uemura, and M.Yamamoto: Appl. Phys. Lett. **90** (2007) 012508.

[7] Y. F. Ding, J. H. Judy, and J. P. Wang: J. Appl. Phys. **97** (2005) 10J117.

[8] Q. L. Lv, J. W. Cai, H. Y. Pan, and B. S. Han: Appl. Phys. Exp. **3** (2010) 093003.

[9] S. Ikeda, K. Miura, H. Yamamoto, K. Mizunuma, H. D. Gan, M. Endo, S. Kanai, J. Hayakawa, F. Matsukura, and H. Ohno: Nat. Mater. **9** (2010) 721.

[10] S. Monso, B. Rodmacq, S. Auffret, G. Casali, F. Fettar, B. Gilles, B. Dieny, and P. Boyer: Appl. Phys. Lett. **80** (2002) 4157.

[11] B. Rodmacq, S. Auffret, B. Dieny, S. Monso, and P. Boyer: J. Appl. Phys. **93** (2003) 7513.

[12] A. Manchon, C. Ducruet, L. Lombard, S. Auffret, B. Rodmacq, B. Dieny, S. Pizzini, J. Vogel, V. Uhlíř, M. Hochstrasser, and G. Panaccione: J. Appl. Phys. **104** (2008) 043914.

[13] Y. Dahmane, S. Auffret, U. Ebels, B. Rodmacq, and B. Dieny: IEEE Trans. Magn. **45** (2008) 3472.

[14] L. E. Nistor, B. Rodmacq, S. Auffret, and B. Dieny: Appl. Phys. Lett. **94** (2009) 012512.

[15] J. H. Jung, S. H. Lim, and S. R. Lee: Appl. Phys. Lett. **96** (2010) 042503.

[16] W. Wang, H. Sukegawa, and K. Inomata: Appl. Phys. Express **3** (2010) 093002.

[17] E. Y. Tsymbal, K. D. Belashchenko, J. P. Velev, S. S. Jaswal, M. van Schilfgaarde, I. I. Oleynik, and D. A. Stewart: Prog. Mater. Sci. **52** (2007) 401.

[18] I. I. Oleinik, E. Yu. Tsymbal, and D. G. Pettifor: Phys. Rev. B **62** (2000) 3952.

[19] K. D. Belashchenko, E. Y. Tsymbal, I. I. Oleinik, and M. van Schilfgaarde: Phys. Rev. B **71** (2005) 224422.

**Figure captions**

Fig. 1. M-H loops for the sample Pt (20)/$Co_2FeAl$ (2)/MgO (0.5) after annealing at different temperatures ($T_{an}$).

Fig. 2. The perpendicular $H_c$ and $H_s$ as functions of $T_{an}$ for the sample Pt (20)/$Co_2FeAl$ (2)/MgO (0.5). The inset shows the MFM image of the 350 $^oC$ annealed sample.

Fig. 3. Co and Fe 2p XPS spectra for the sample of Pt (20)/$Co_2FeAl$ (2)/MgO (0.5) annealed at 350 $^oC$.

Fig. 4. $M_s$ and the effective thickness of $Co_2FeAl$ ($t_{eff}$) as a function of $t_{MgO}$ for the as-deposited Pt (20)/$Co_2FeAl$ (2)/MgO ($t_{MgO}$) samples. Inset: Perpendicular M-H loop for the as-deposited Pt (5)/$Co_2FeAl$ (2)/Pt (4) structure.

Fig. 5. Perpendicular M-H loop for the 250 $^oC$ annealed sample of Pt (20)/$Co_2FeAl$ (2.5)/MgO (0.5). The inset shows M-H loops for the as-deposited sample.

Fig. 6. Variations of the perpendicular $H_s$ with $t_{CFA}$ for the samples of Pt (20)/$Co_2FeAl$ ($t_{CFA}$)/MgO (0.5) in the as-deposited and 350 $^oC$ annealed states, respectively. The inset shows M-H loops for the 350 $^oC$ annealed sample with $t_{CFA}$=2.5 nm.

Fig. 1 (X.Q.Li et al.):

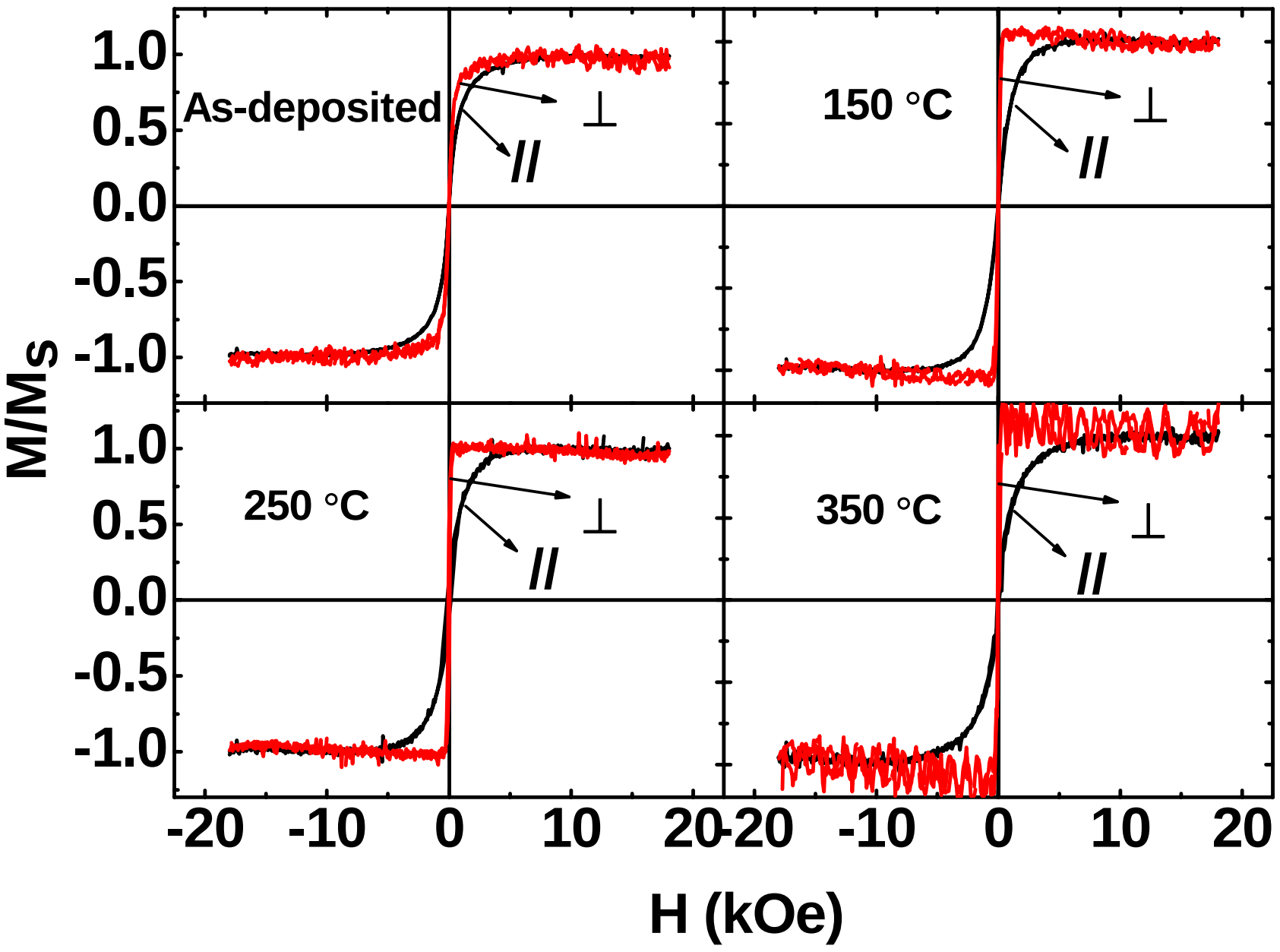

Fig. 2 (X.Q.Li et al.):

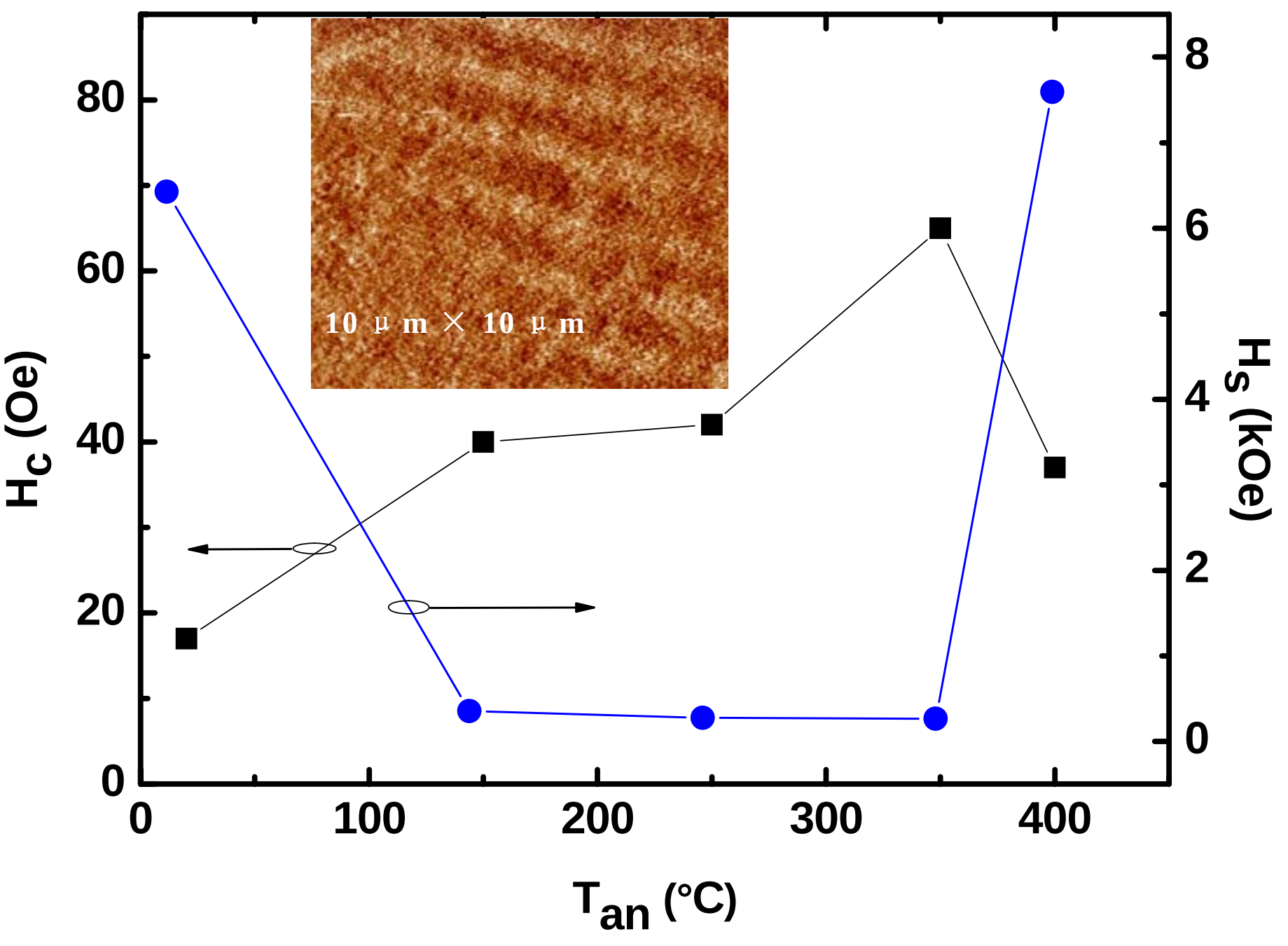

Fig. 3 (X.Q.Li et al.):

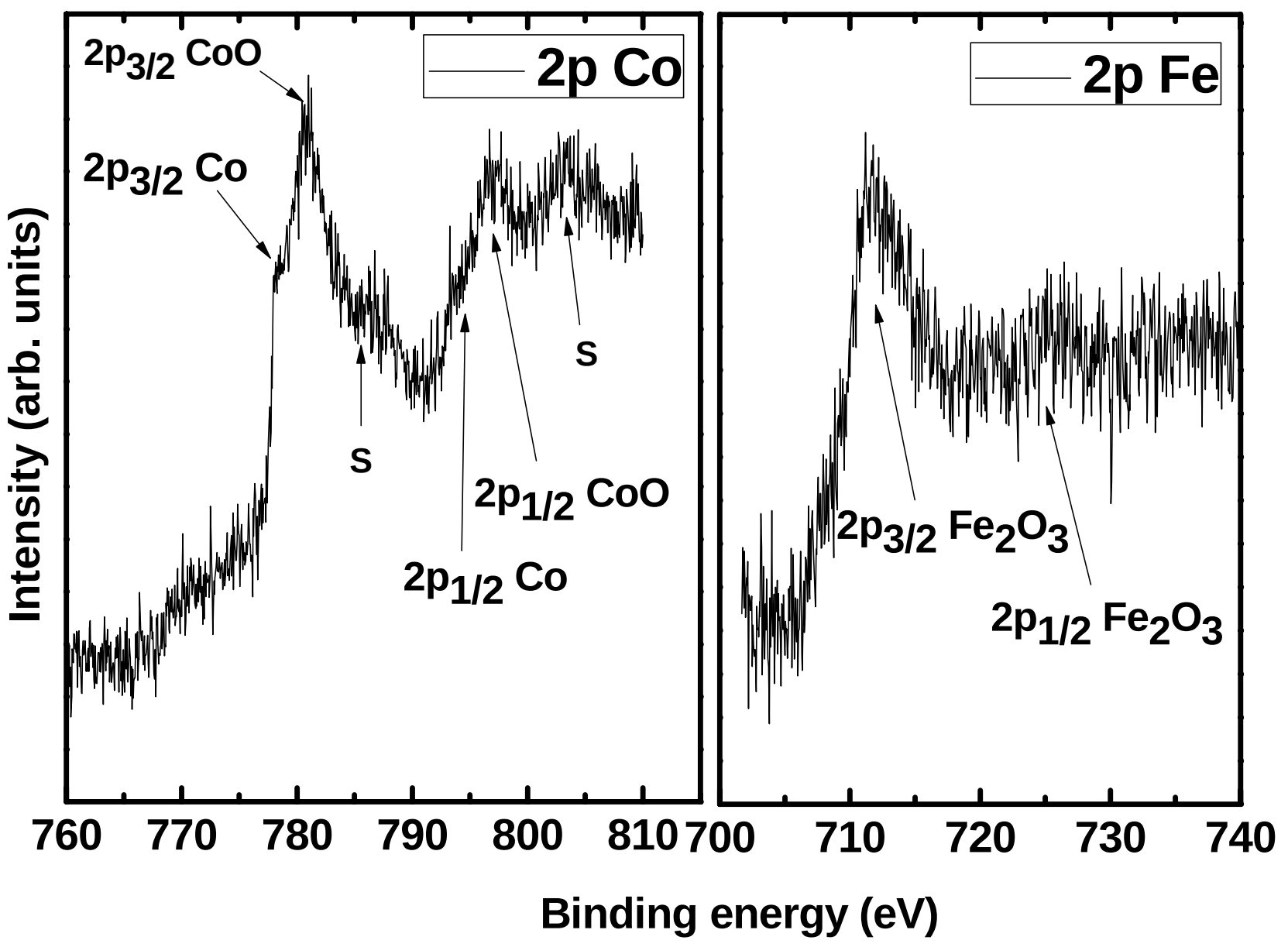

Fig. 4 (X.Q.Li et al.):

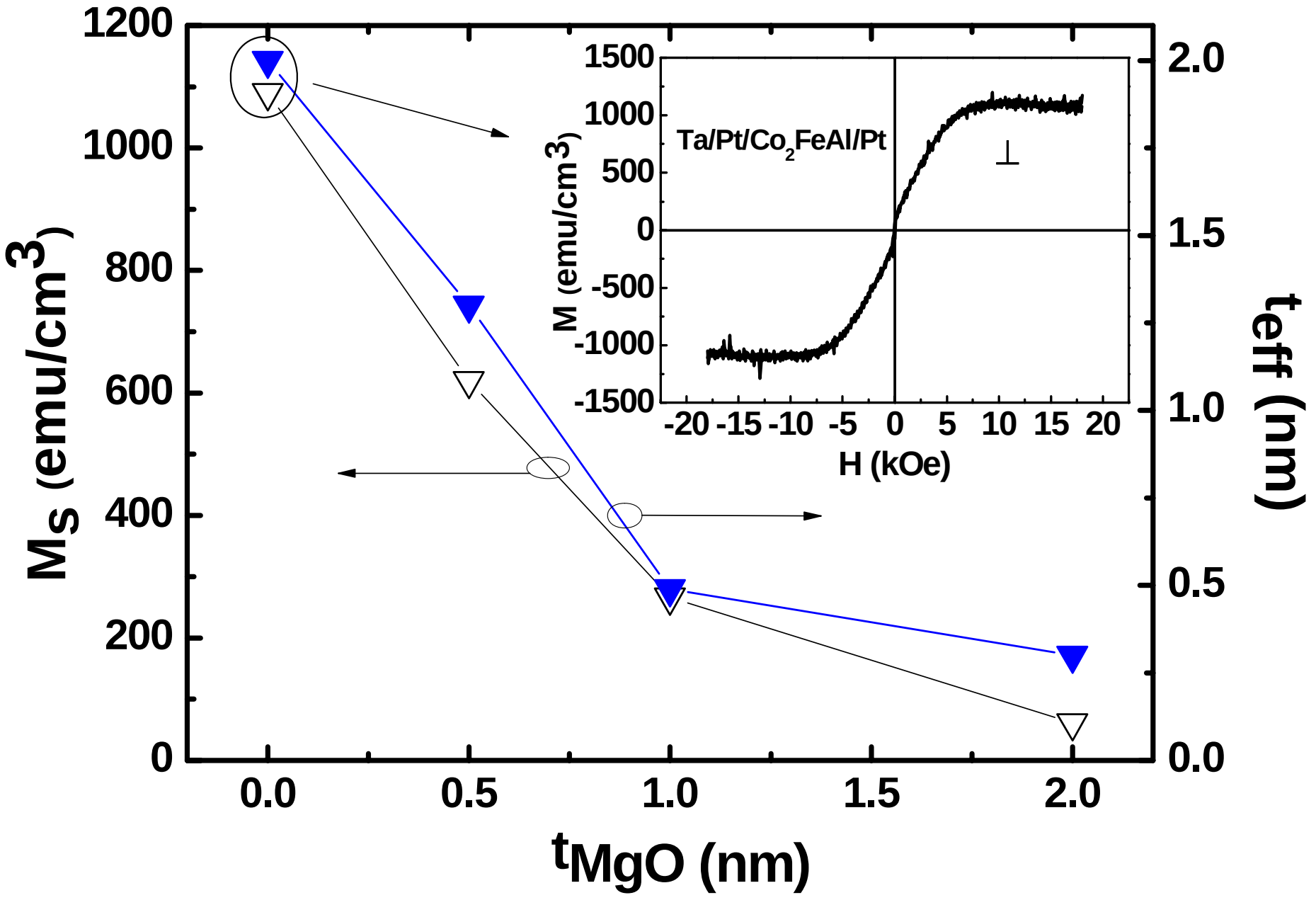

Fig. 5 (X.Q.Li et al.):

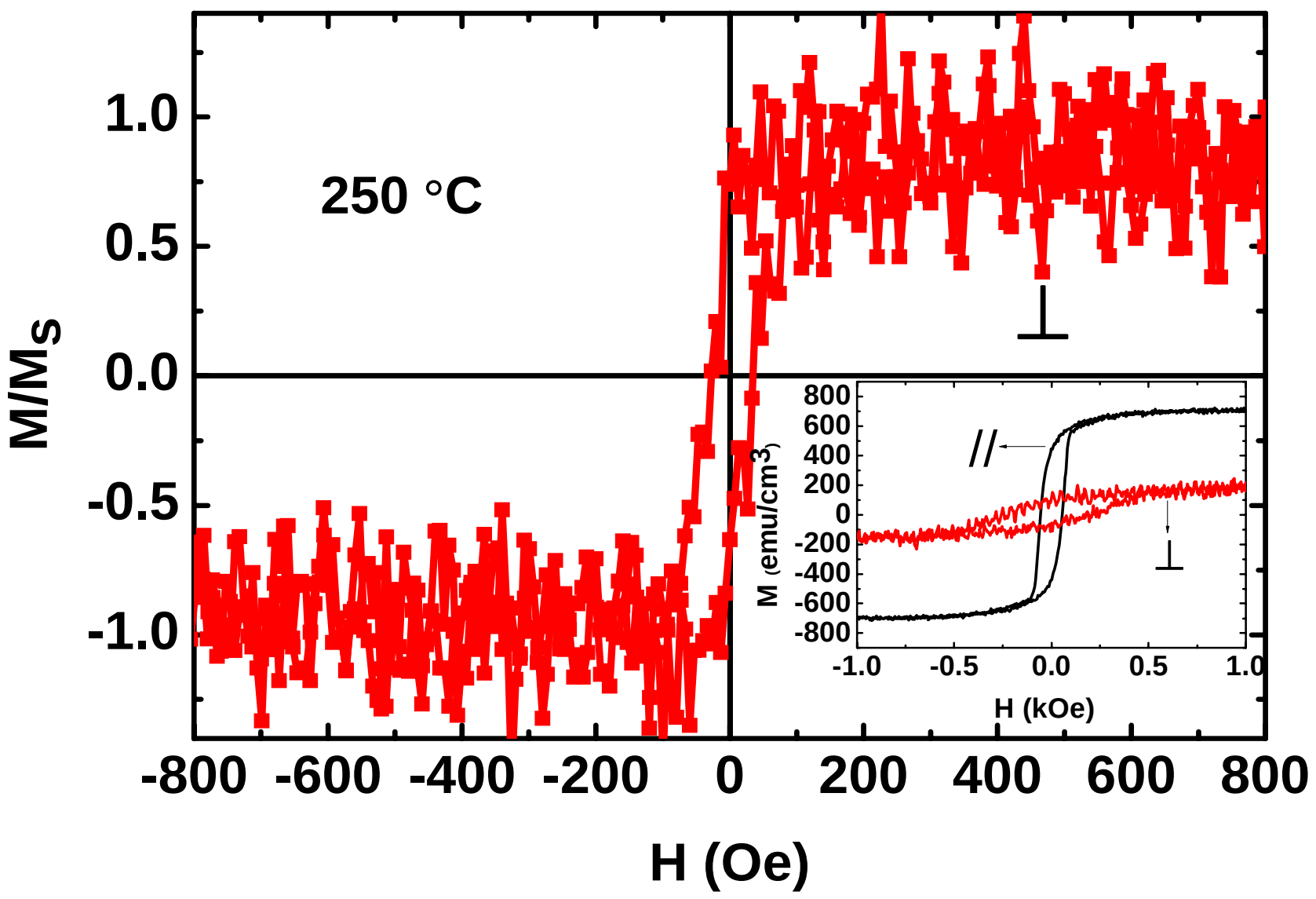

Fig. 6 (X.Q.Li et al.):

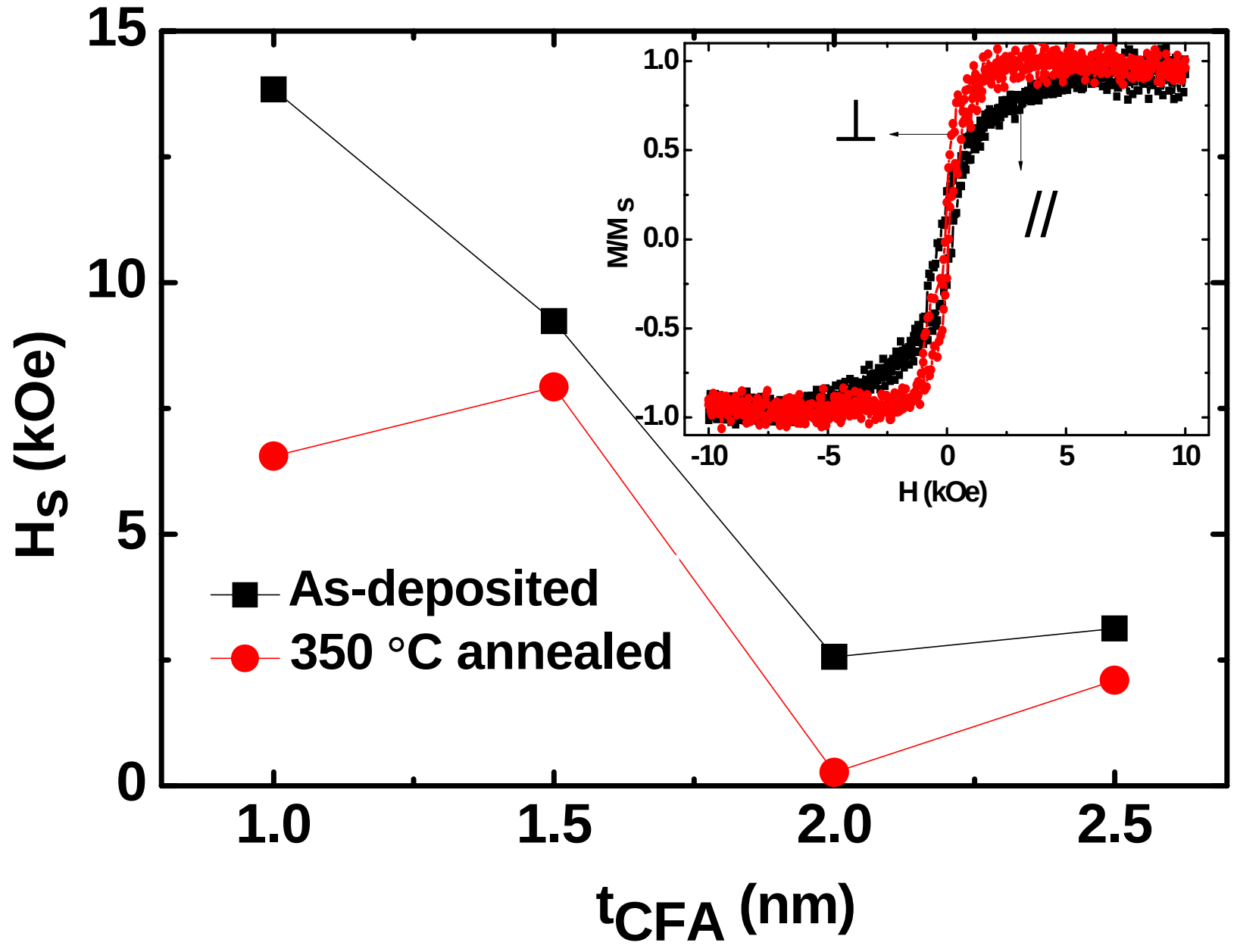